\documentstyle[12pt]{article}
\begin{document}
\title{INTEGRABLE COUPLED KdV SYSTEMS}
\author{ Metin G{\" u}rses\\
{\small Department of Mathematics, Faculty of Sciences}\\
{\small Bilkent University, 06533 Ankara - Turkey}\\
{\small and}\\
Atalay Karasu\\
{\small Department of Physics , Faculty of Arts and  Sciences}\\
{\small Middle East Technical University , 06531 Ankara-Turkey}}
\date{}
\begin{titlepage}
\maketitle
\begin{center}
PACS ( 02.30.Jr, 02.90.+p, 11.30 Na, 11.30. Ly.)
\end{center}

\begin{abstract}
We give the conditions for a system of $N$- coupled Korteweg de Vries(KdV)
type of equations to be integrable. Recursion operators of each subclasses
are also given. All examples for $N=2$ are explicitly given.
\end{abstract}
\end{titlepage}

{\bf{I. Introduction}} \\
In \cite{MET} we gave an extension of the recently proposed Svinolupov
Jordan KdV \cite{SVI}\cite{SV1} systems to a class of integrable
multicomponent KdV systems and gave their recursion operators. This class
is known as the degenerate subclass of the KdV system.
In this work we will extend it to a more general KdV type of
system equations containing both the degenerate and non-degenerate cases.
This is a major step towards the complete classification
of KdV systems. In addition we give a new extension of such system of
equations.

Let us consider a system of $N$ nonlinear equations of the form

\begin{equation}
 q^{i}_{t}=b^{i}_{j}q^{j}_{xxx}+s^{i}_{j k}q^{j}q^{k}_{x} , \label{a0}
\end{equation}

\noindent
where $i,j,k=1,2,...,N$ ,
$q^{i}$ are functions depending on the variables $x$ , ${t}$\, , \,
and  $b^{i}_{j}$\,\, ,\,\,$S^{i}_{jk}$
are constants. The purpose of this work is to find the conditions
on these constants so that the equations in (\ref{a0}) are integrable.
In general the existence of infinitely many conserved quantities
is admited as the definition of integrability. This implies the
existence of infinitely many generalised symmetries .
In this work we assume the following definition for integrability:

\noindent
{\bf Definition}: A system of equation is said to be integrable
if it admits a recursion operator.

\noindent
The recursion operator (if it exists) of the system of equations
given in (\ref{a0}) in general, may take a very complicated form.
Let the highest powers of the operators $D$ and $D^{-1}$ be
respectively defined by $m$=degree of $R$ and $n$=order non-locality of $R$.
In this work we are interested in a subclass of equations admitting
a recursion operator with $m=2$ and $n=1$. Namely it is of the form

\begin{equation}
R^{i}_{j}=b^{i}_{j}\,D^{2}+a^{i}_{jk}\,q^{k}+c^{i}_{jk}\,q^{k}_{x}\,D^{-1}
 \label{a1}
\end{equation}

\noindent
where $D$ is the total $x-$ derivative , $D^{-1}$ is the inverse operator
and $a^{i}_{jk}$ , $c^{i}_{jk}$ are constants with

\begin{equation}
 s^{i}_{j\,k}=a^{i}_{k\,j}+c^{i}_{j\,k}. \label{a3}
 \end{equation}

\noindent
Before starting to the classification  of (\ref{a0}) we recall
a few fundamental properties of the recursion operator.
An operator $R^{i}_{j}$ is a recursion operator if it satisfies
the following equation

\begin{equation}
R^{i}_{j , t}= \,F^{\prime \, i}_{k}\,\,R^{k}_{j}\,\,-\,\,R^{i}_{k}\,\,
F^{\prime \,k}_{j} ,    \label{b3}
\end{equation}

\noindent
where $F^{\prime \, i}_{k}$ is the Fr{\'e}chet derivative of the
system (\ref{a0}) which is given by

\begin{equation}
\sigma^{i}_{t}=F^{\prime \,i}_{j}\, \sigma^{j} ,\label{b1}
\end{equation}

\noindent
where $\sigma^{i}$'s are called the symmetries of the system (\ref{a0}).
The condition (\ref{a3}) implies that equation (\ref{a0}) is itself
is assumed to be in the family of the hierarchy of equations (or flows)

$$q^{i}_{\tau_{n}}=\sigma^{i}_{n}$$

\noindent
where for all $n=0,1,2,...$ ,\,\,$\sigma^{i}_{n}$ denotes the symmetries
of the integrable KdV system (\ref{a0}). For instance for $n=0,1$ we have
respectively the classical symmetries $\sigma^{i}_{0}=q^{i}_{,x}$ and
$\sigma^{i}_{1}=q^{i}_{,t}$.

Eq.(\ref{b1}) is called the symmetry equation of (\ref{a0}) with

\begin{equation}
F^{\prime \,i}_{j}=b^{i}_{j}\,D^{3}+s^{i}_{jk}\,q^{k}_{x}+
s^{i}_{kj}\,q^{k}\,D .
\end{equation}

\noindent
Recursion operators are defined as operators mapping symmetries to
symmetries , i.e

\begin{equation}
R^{i}_{j}\, \sigma^{j}= \lambda \, \sigma^{i} ,  \label{b2}
\end{equation}

\noindent
where $\lambda$ is an arbitrary constant. Equations (\ref{b1}) and
(\ref{b2}) imply (\ref{b3}). It is the equation (\ref{b3}) which
determines the constants
$a^{i}_{jk}$ , $c^{i}_{j\,k}$  in terms of
$b^{i}_{j}$ and $s^{i}_{jk}$'s. The same equation (\ref{b3}) brings
severe constraints on  $b^{i}_{j}$ and $s^{i}_{jk}$.

We shall obtain a classification of (\ref{a0}) based on the
matrix $b^{i}_{j}$.

\noindent
({\bf I}). $det(b^{i}_{j}) = 0$ ,\\
({\bf II}). $det(b^{i}_{j})\not= 0$,

\noindent
and also we  divide the classification procedure,for each class,into two
parts where $s^{i}_{jk}=s^{i}_{kj}$ and $s^{i}_{jk}\not=s^{i}_{kj}$.
For the system of equations admitting a recursion operator we have the
following proposition

\noindent
{\bf Proposition 1}: Let $q^{i}(t,x)$ be functions of $t$ and $x$ satisfying
the $N$ KdV equations (\ref{a0}) and admitting a recursion operator
$R^{i}_{j}$ in (\ref{a1}) then the constants $b^{i}_{j}\, , \,
s^{i}_{jk}\, , \, a^{i}_{jk}\, , \, c^{i}_{jk}$ satisfy (in addition to the
(\ref{a3})) the following relations

\begin{eqnarray}
&&b^{k}_{l} c^{i}_{j\,k}-b^{i}_{k} c^{k}_{j\,l}=0, \label{b4}
\\
&&b^{k}_{l} a^{i}_{j\,k}-b^{i}_{k} (a^{i}_{j\,l}+3 c^{k}_{j\,l}
- s^{k}_{j\,l})=0,\label{b5}
\\
&& b^{i}_{k}(3 a^{k}_{j\,l}+3 c^{k}_{j\,l}
- 2 s^{k}_{j\,l}- s^{k}_{l\,j})=0,\label{b6}
\\
&& c^{i}_{j\,k} s^{k}_{l\,m}- s^{i}_{l\,k} c^{k}_{j\,m}=0,
  \label{b7}
\\
&& c^{i}_{j\,k} s^{k}_{l\,m}+ c^{i}_{j\,k}s^{k}_{m\,l}
-c^{k}_{j\,m}s^{i}_{k\,l}- c^{k}_{j\,l}s^{i}_{k\,m}=0,\label{b8}
\\
&& a^{i}_{j\,k} s^{k}_{l\,m}- s^{i}_{k\,m}a^{k}_{j\,l}
-s^{i}_{l\,k}a^{k}_{j\,m}- s^{i}_{l\,k}c^{k}_{j\,m}
+s^{k}_{j\,l}c^{i}_{k\,m}+ a^{i}_{k\,l}s^{k}_{j\,m}
=0, \label{b10}
\\
&& c^{i}_{k\,m} (s^{k}_{p\,j}- s^{k}_{j\,p})=0. \label{b11}
 \end{eqnarray}

\noindent
Now we will discuss the problem of classifying integrable system of
equations (\ref{a0}) for the two exclusive cases depending upon the
matrix $b^{i}_{j}$.

{\bf{II. Classification for the class $det(b^{i}_{j})= 0$.}} \\
In this subclass we assume the rank of the matrix $b^{i}_{j}$ as $N-1$.
Investigation of the subclasses for other ranks of matrix $b$ can be
done similarly.
For this case we may take $b^{i}_{j}=\delta^{i}_{j}-k^{i}k_{j}$
where $k_{i}$ is a unit vector , $k^{i}\,k_{i}=1$. In this work
we use the Einstein convention , i.e., repeated indices are summed up
from 1 to $N$ . We find the following solution
of (\ref{b4}-\ref{b11}) for the parameters $a^{i}_{j\,k}$ and
 $c^{i}_{j\,k}$ for all $N$

\noindent
{\bf Proposition 2}: Let $k^{i}$ be a constant unit vector and
$b^{i}_{j}= \delta^{i}_{j}-k^{i}\,k_{j}$ then the complete solution of
the equations (\ref{b4}-\ref{b11}) are given by

\begin{eqnarray}
&&a^{i}_{l\,j}={2 \over 3}\,s^{i}_{j\,l}+
{1 \over 3}\,[k^{i}\,(k_{j}\,a_{l}-2k_{l}\,n_{j})
 +(-a\,k^{i}+b^{i})k_{l}k_{j}] , \nonumber \\
&&c^{i}_{j\,l}= {1 \over 3}s^{i}_{l\,j}-
{1 \over 3}\,[k^{i}(k_{l}\,a_{j}+ k_{j}\,n_{l})+
(-a\,k^{i}+b^{i})k_{l}k_{j}]+k^{i}k_{l}n_{j}, \label{b12}
\end{eqnarray}

\noindent
where

\begin{eqnarray}
&&n_{j}=k^{k}k_{i}s^{i}_{j\,k}-a\,k_{j},  \nonumber \\
&&a_{j}=k^{k}k_{i}s^{i}_{k\,j}-n\,k_{j},   \nonumber \\
&&b^{i}=k^{l}k^{k}s^{i}_{l\,k}-n\,k^{i}.  \label{b13}
\end{eqnarray}

\noindent
where $a=k^{n}a_{n}$, $n=k^{i}n_{i}$ and $s^{i}_{jk}$'s are
subject to satisfy the following

\begin{eqnarray}
c^{i}_{jk}\,s^{k}_{lm}=c^{k}_{jm}\,s^{i}_{lk}, \\
s^{i}_{jk}-s^{i}_{kj}=k^{i}\,[k_{j}(a_{k}-n_{k})-k_{k}(a_{j}-n_{j})], \\
k_{n}\,s^{n}_{lj}=n_{l}k_{j}+k_{l}a_{j}, \\
k^{n}\,s^{i}_{jn}=k^{i}n_{j}+b^{i}k_{j}, \\
k^{n}\,s^{i}_{nj}=(n-a)k^{i}k_{j}+k^{i}a_{j}+b^{i}k_{j}, \\
n\,[(a_{i}-n_{i})\,k_{j}-(a_{j}-n_{j})\,k_{i}]=0, \label{as1} \\
a_{i}=\rho\,n_{i}+a\,k_{i},
\end{eqnarray}

\noindent
where $\rho$ is a constant.
At this point we will discuss the classification procedure with respect
to the symbol $s^{i}_{jk}$ whether it is symmetric or non-symmetric
with respect to its lower indices.

\noindent
{\bf (A). The symmetric case, $s^{i}_{jk}=s^{i}_{kj}$ :}
Among the constraints listed in proposition 2 the one given in
(\ref{as1}) implies
that $s^{i}_{jk}$'s are symmetric if and only if $a_{i}-n_{i}=\alpha\,k_{i}$
where $\alpha=n-a$.
There are two distinct cases depending on whether $n=0$ or  $n \not=0$.
We shall give these two sub-cases as corollaries of the previous
proposition.

\noindent
{\bf Corollary 1}:
Let  $s^{i}_{jk}=s^{i}_{kj}$ and $n=0$. Then we have the following
solution for all $N$.

\begin{eqnarray}
&&a^{i}_{k\,j}={2 \over 3}\,s^{i}_{j\,k}+
{1 \over 3}\,[k^{i}\,(k_{j}\,a_{k}-2k_{k}\,n_{j})
 +k_{k}\,k_{j}\,b^{i}] , \nonumber \\
&&c^{i}_{j\,k}= {1 \over 3}s^{i}_{j\,k}-
{1 \over 3}\,[k^{i}(k_{j}\,a_{k}- 2 k_{k}\,n_{j})+k_{k}\,k_{j}\,b^{i}],
\label{b14}
\end{eqnarray}

\noindent
where $a=0$ , $\rho=1$ and

\begin{equation}
a_{l}=n_{l},~
n_{l}=k_{i}k^{j}s^{i}_{l\,j},~
b^{i}=k^{j}k^{l}s^{i}_{l\,j}.
\end{equation}

\noindent
The vector $k^{i}$ and $s^{i}_{jk}$ are not arbitrary , they satisfy
the following constraints

\begin{eqnarray}
s^{i}_{jk} s^{k}_{l\,m}-s^{i}_{lk} s^{k}_{jm}=-2\,(k_{j}\,n_{l}-
k_{l}\,n_{j})\,(-k^{i}\,n_{m}+b^{i}\,k_{m}) ,\nonumber \\
k_{n}\,s^{n}_{lj}=n_{l}k_{j}+k_{l}\,n_{j},\nonumber \\
k^{n}\,s^{i}_{jn}=k^{i}n_{j}+b^{i}k_{j},\label{b15} .
\end{eqnarray}

\noindent
As an illustration we give an example to this case \cite{MET}.
A particular solution of the equations
(\ref{b15}) for  $N=2$

\begin{eqnarray}
&&b^{i}_{j}=\delta^{i}_{j}-y^{i}\,y_{j}=x^{i}\,x_{j} ,\nonumber  \\
&&s^{i}_{j\,k}={3 \over 2}\alpha_{1}\, x^{i}\,x_{j}\,x_{k}
+\alpha_{2} x^{i}\,y_{j}\,y_{k}+
{\alpha_{1} \over 2}
 y^{i}\,(y_{j}\,x_{k}+ y_{k}\,x_{j}) ,
\end{eqnarray}

\noindent
where $i,j=1,2$ and

\begin{equation}
 x^{i}=\delta^{i}_{1} ~~,~~ y^{i}= \delta^{i}_{2} ,
\end{equation}

\noindent
and

\noindent
\begin{equation}
k_{i}=y_{i}~ ,~ n_{i}={1 \over 2}\alpha_{1}x_{i}~, ~b^{i}=\alpha_{2}x^{i}.
\end{equation}

\noindent
Constants $a^{i}_{jk}$ and $c^{i}_{jk}$ appearing in the recursion
operator are given by

\begin{eqnarray}
&&a^{i}_{j\,k}=\alpha_{1} \,x^{i}\,x_{j}\,x_{k}
+\alpha_{2}\, x^{i}\,y_{j}\,y_{k}
+{\alpha_{1} \over 2}\,y^{i}\,x_{j}\, y_{k}, \nonumber\\
&&c^{i}_{j\,k}={\alpha_{1}\over 2} \,x^{i}\,x_{j}\,x_{k}+
{\alpha_{1} \over 2} y^{i}\,x_{j}\,y_{k} .
\end{eqnarray}

\noindent
Taking $\alpha_{1}=2$ and  $\alpha_{2}=1$ (without loss of generality)
we obtain the following coupled system

\begin{eqnarray}
 &&u_{t}=u_{xxx}+3uu_{x}+vv_{x} , \nonumber \\
 &&v_{t}=(uv)_{x}.     \label{ito}
\end{eqnarray}

\noindent
The above system was first introduced by Ito \cite{ITO} and the
biHamiltonian structure has been studied by Olver and Rosenau \cite{OLV1}.
The recursion operator of this system is given by

\begin{equation}
R=
\: \left(
\begin{array}{cc}
D^{2}+2\,u +u_{x}D^{-1}
 & v \\
v+v_{x}D^{-1} &
0
\end{array} \; \; \right)\;.
\label{rec}
\end{equation}

\noindent
In \cite{MET} we have another example for $N=3$.

\noindent
For the case $n \not=0$ for all $N$ we have

\noindent
{\bf Corollary 2:} Let $s^{i}_{jk} = s^{i}_{kj}$ , $n \not =0$
and $\rho=0$ then
the solution given in proposition 2 reduces to

\begin{eqnarray}
&&a^{i}_{l\,j}={2 \over 3}\,s^{i}_{j\,l}+
{1 \over 3} \, (a-2\,n) \, k^{i}\,k_{j}\,k_{l} , \nonumber \\
&&c^{i}_{j\,l}= {1 \over 3}s^{i}_{l\,j}-
{1 \over 3}\, (a-2\,n)\, k^{i}\,k_{j}\,k_{l} \label{b16}
\end{eqnarray}

\noindent
where

\begin{equation}
n_{j}=n\,k_{j},~
a_{j}=a\,k_{j},~
b^{i}=a\,k^{i},
\end{equation}

\noindent
and the constraint equations

\begin{eqnarray}
s^{i}_{kj} s^{k}_{m\,l}-s^{k}_{mj} s^{i}_{kl}=0 \nonumber \\
k_{n}\,s^{n}_{lj}=(a+n)\,k_{l}k_{j} \nonumber \\
k^{n}\,s^{i}_{jn}=(a+n)\,k^{i}k_{j} \label{b17}
\end{eqnarray}

\noindent
For this case we point out that solution of (\ref{b16}) and (\ref{b17})
gives  decoupled systems.

\noindent
{\bf(B). Non-symmetric case $s^{i}_{jk}\not=s^{i}_{kj}$:} In this case
the constraints in proposition 1 , in particular (\ref{as1})
implies that we must have $n=0$. In this case we have the following
expressions for $a^{i}_{l\,j}$ and $c^{i}_{l\,j}$ for all $N$

\begin{eqnarray}
&&a^{i}_{l\,j}={2 \over 3}\,s^{i}_{j\,l}+
{1 \over 3}\,[k^{i}\,(k_{j}\,a_{l}-2k_{l}\,n_{j})
 +(-a\,k^{i}+b^{i})k_{l}k_{j}] , \nonumber \\
&&c^{i}_{j\,l}= {1 \over 3}s^{i}_{l\,j}-
{1 \over 3}\,[k^{i}(k_{l}\,a_{j}+ k_{j}\,n_{l})+(-a
\,k^{i}+b^{i})k_{l}k_{j}]+k^{i}k_{l}n_{j}, \label{b18}
\end{eqnarray}

\noindent
where

\begin{eqnarray}
&&n_{j}=k^{k}k_{i}s^{i}_{j\,k}-ak_{j},  \nonumber \\
&&a_{j}=k^{k}k_{i}s^{i}_{k\,j},   \nonumber \\
&&b^{i}=k^{l}k^{k}s^{i}_{l\,k}.
\end{eqnarray}

\noindent
and the constraint equations among the parameters are

\begin{eqnarray}
&&c^{i}_{jk} c^{k}_{l\,m}-c^{i}_{lk} c^{k}_{jm}=0  \\
&&(a^{i}_{jk}- c^{i}_{j\,k})s^{k}_{lm}+
(c^{i}_{km}- a^{i}_{km})s^{k}_{lj}
+(s^{i}_{mk}- s^{i}_{km})a^{k}_{jl} \nonumber \\
&+&(s^{k}_{jm}- s^{k}_{mj})a^{i}_{kl}=0.
\end{eqnarray}

\noindent
For $N=2$ we will give an example.
Constants $a^{i}_{jk}$ and $c^{i}_{jk}$ appearing in the recursion
operator are given by

\begin{eqnarray}
&&a^{i}_{j\,k}=\alpha_{1} \,x^{i}\,x_{j}\,x_{k}
+\alpha_{2}\, x^{i}\,y_{j}\,y_{k}
+\alpha_{3} \,y^{i}\,x_{j}\, y_{k}, \nonumber\\
&&c^{i}_{j\,k}={\alpha_{1}\over 2} \,x^{i}\,x_{j}\,x_{k}+
\alpha_{1} \,y^{i}\,x_{j}\,y_{k} .
\end{eqnarray}
\noindent
where  $\alpha_{1}$,$ \alpha_{2}$ and  $\alpha_{3}$ are arbitrary
constants and
\begin{equation}
k_{i}=y_{i}~ ,~ n_{i}=\alpha_{3}x_{i}~, ~b^{i}=\alpha_{2}x^{i},
a_{i}=\alpha_{1}x_{i}.
\end{equation}

\noindent
We obtain the following coupled system

\begin{eqnarray}
 &&u_{t}=u_{xxx}+3 \alpha_{1} u u_{x} ,\nonumber \\
 &&v_{t}=\alpha_{3} u_{x} v+\alpha_{1} u v_{x}.\label{b19}
\end{eqnarray}

\noindent
which is equivalent to the symmetrically coupled KdV system \cite{FUC}

\begin{eqnarray}
 &&u_{t}=u_{xxx}+v_{xxx}+6uu_{x}+4uv_{x}+2u_{x}v,\nonumber \\
 &&u_{t}=u_{xxx}+v_{xxx}+6vv_{x}+4vu_{x}+2v_{x}u
\end{eqnarray}

\noindent
and the recursion operator for this integrable system of equations
 (\ref{b19}) is

\begin{equation}
R=
\: \left(
\begin{array}{cc}
D^{2}+\alpha_{1}( 2u + u_{x}D^{-1})
 & 0 \\
\alpha_{3} v+\alpha_{1} v_{x}D^{-1} &
0
\end{array} \; \; \right)\;.
\end{equation}

{\bf{III. Classification for the case $det(b^{i}_{j})\not = 0$.}} \\
As in the degenerate case $det(b^{i}_{j})=0$ we have two sub-cases,
symmetric and non-symmetric, before these we have the following
proposition

\noindent
{\bf Proposition 3}: Let $det(b^{i}_{j}) \not =0$ then the
solution of equations given proposition 1 is given as follows

\begin{eqnarray}
&&a^{i}_{j\,l}={2 \over 9}\,(s^{i}_{l\,j}+2s^{i}_{j\,l})
-{1 \over 9}\,C^{k}_{l}b^{i}_{m}K^{m}_{j\,k}  \nonumber    \\
&&c^{i}_{l\,j}={1 \over 9}\,(7s^{i}_{l\,j}-4s^{i}_{j\,l})
+{1 \over 9}\,C^{k}_{l}b^{i}_{m}K^{m}_{j\,k}. \label{b20}
\end{eqnarray}

\noindent
where

\begin{equation}
K^{i}_{lj}=s^{i}_{l\,j}-s^{i}_{j\,l}
\end{equation}

\noindent
and the constraint equations

\begin{eqnarray}
&&b^{k}_{l} c^{i}_{j\,k}-b^{i}_{k} c^{k}_{j\,l}=0, \label{b21}   \\
&&5C^{m}_{i}K^{i}_{rj}-C^{l}_{r}K^{m}_{lj} -C^{l}_{j}K^{m}_{rl}
=0 \label{b22} \\
&&c^{k}_{jm}K^{i}_{lk}+c^{k}_{jl}K^{i}_{mk}=0  \label{b23}    \\
&&c^{i}_{jk}s^{k}_{lm}-c^{k}_{jm}s^{i}_{lk}=0 \label{b24}\\
&&K^{k}_{lj}c^{i}_{km}=0, \label{b25}\\
&&(a^{i}_{jk}- c^{i}_{j\,k})s^{k}_{lm}+
(c^{i}_{km}- a^{i}_{km})s^{k}_{lj}
+K^{i}_{mk}a^{k}_{jl}+
K^{k}_{jm}a^{i}_{kl}=0. \label{b26}
\end{eqnarray}
\noindent

\noindent
where $C^{i}_{r}$ is the inverse of $b^{r}_{i}$.

\noindent
{\bf (A). Non-symmetric case  $s^{i}_{jk}\not=s^{i}_{kj}$ :}
Eqs. (\ref{b21}-\ref{b26}) define an over-determined system for the
components of $s^{i}_{jk}$ . Any solution of this system leads to
the determination of the parameters $a^{i}_{j\,l}$ and  $c^{i}_{j\,l}$
by (\ref{b20}).

As an example we give the following,for $ N=2$,coupled system

\begin{eqnarray}
 &&v_{t}=av_{xxx}+2bvv_{x},\nonumber \\
 &&u_{t}=4au_{xxx}+2bu_{x}v+buv_{x} ,\label{b27}
\end{eqnarray}

\noindent
where $a$ and $b$ are arbitrary constants.
This system ,under a change of variables,
is equivalent to the KdV equation with the time evolution part of
its Lax equation \cite{ABL}. The recursion operator of the system
(\ref{b27}) is

\begin{equation}
R=
\: \left(
\begin{array}{cc}
{4 \over 3}(3aD^{2}+bv)
 & {b \over 3}(3u +2 u_{x}D^{-1})\\
0 &
{1 \over 3}(3 a D^{2}+4 b v+2 b v_{x}D^{-1})
\end{array} \; \; \right)\;.
\end{equation}

\noindent
Hence the KdV equation coupled to time evolution part of its
Lax-pair is integrable and its recursion operator is given above.
This is the only new example for $N=2$ system.

\noindent
{\bf (B). Symmetric case $s^{i}_{jk}=s^{i}_{kj}$ :}
When the symbol $s^{i}_{j\,k}$ is symmetric with
respect to sub-indices the parameters $K^{i}_{j\,k}$ vanish.
Then the  equations (\ref{b20}-\ref{b26}) reduce to

\begin{equation}
a^{i}_{j\,k}={2 \over 3}s^{i}_{j\,k}\,\,\,\, ,
 c^{i}_{j\,k}={1 \over 3 }s^{i}_{j\,k} , \\
\end{equation}

where the parameters $b^{i}_{k}$ and $s^{i}_{j\,k}$ satisfy
\begin{eqnarray}
&&b^{k}_{l} s^{i}_{j\,k}-b^{i}_{k} s^{k}_{j\,l}=0 ,\\
&&s^{i}_{j\,k}s^{k}_{l\,m}- s^{i}_{l\,k}s^{k}_{j\,m}=0.
\end{eqnarray}

\noindent
We shall not study this class in detail , because in \cite{MET}
some examples of this class are given for $N=2$. Here we give
another example which correspond to the perturbation expansion
of the KdV equation. Let $q^{i}=\delta^{i}\,u$ where $i=0,1,2,...,N$
and $u$ satisfies the KdV equation $u_{,t}=u_{,xxx}+6\,u\,u_{,x}$.
The $q^{i}$'s satisfy a system of KdV equations which belong to this class

\begin{eqnarray} \displaystyle
q^{0}_{,t}=q^{0}_{,xxx}+6\,q^{0}\,q^{0}_{,x},\\
q^{1}_{,t}=q^{1}_{,xxx}+6\,(q^{0}\,q^{1})_{,x},\\
q^{2}_{,t}=q^{2}_{,xxx}+6\,[(q^{1})^{2}+q^{0}\,q^{2}]_{,x},\\
................\\
q^{N}_{,t}=q^{N}_{,xxx}+3\,\sum_{i=1}^{N} [\delta^{i}(q^{0})^{2}]_{,x}
\end{eqnarray}

{\bf{IV. Fokas-Liu Extension}} \\
The classification of the KdV system given in this work with respect
to the symmetries  can be easily extended to the following simple
modification of (\ref{a0})

\begin{equation}
 q^{i}_{t}=b^{i}_{j}q^{j}_{xxx}+s^{i}_{j k}q^{j}q^{k}_{x} +
 \chi^{i}_{j}\,q^{j}_{,x}, \label{a5}
\end{equation}

\noindent
where $\chi^{i}_{j}$'s are arbitrary constants. Eq.(\ref{a5}) without the
last term will be called the principle part of that equation. Hence the
equation (\ref{a0}) we have studied so far is the principle part of its
modification (\ref{a5}). We assume the existence of a
recursion operator corresponding to the above system in the form

\begin{equation}
R^{i}_{j}=b^{i}_{j}\,D^{2}+a^{i}_{jk}\,q^{k}+c^{i}_{jk}\,q^{k}_{x}\,D^{-1}
+w^{i}_{j} \label{a6}
\end{equation}

\noindent
where $w^{i}_{j}$'s are constants. We have the following proposition
corresponding to  the integrability of the above system

\noindent
{\bf Proposition 4}: The operator given in (\ref{a6}) is the recursion
operator of the KdV system (\ref{a5}) if in addition to the equations
listed in proposition 1 (\ref{b4}-\ref{b11}) the following constraints
on the constants $\chi^{i}_{j}$ and $w^{i}_{j}$ are satisfied

\begin{eqnarray}
\chi^{k}_{l}\,a^{i}_{jk}-\chi^{i}_{k}\,a^{k}_{jl}-\chi^{i}_{k}\,c^{k}_{jl}
+\chi^{k}_{j}\,c^{i}_{kl}-w^{k}_{j}\,s^{i}_{kl}+w^{i}_{k}\,s^{k}_{jl}=0  ,
\label{c1}\\
\chi^{k}_{l}\,c^{i}_{jk}-\chi^{i}_{k}\,c^{k}_{jl}=0  ,\label{c2} \\
\chi^{i}_{k}\,w^{k}_{j}-\chi^{k}_{j}\,w^{i}_{k}=0  , \label{c3} \\
(\chi^{k}_{j}-w^{k}_{j})\,b^{i}_{k}-(\chi^{i}_{k}-w^{i}_{k})\,b^{k}_{j}=0  ,
\label{c4} \\
\chi^{k}_{j}\,a^{i}_{kl}-\chi^{i}_{k}\,a^{k}_{jl}+w^{i}_{k}\,s^{k}_{lj}-
w^{k}_{j}\,s^{i}_{lk}=0 . \label{c5}
\end{eqnarray}

\noindent
Since the constraints (\ref{b4}-\ref{b11}) are enough to determine
the coefficients $a^{i}_{jk}$
and $c^{i}_{jk}$ with some constraints on the given constants $b^{i}_{j}$
and $s^{i}_{jk}$'s we have the following corollary of the above proposition

\noindent
{\bf Corollary 1:} The KdV system in (\ref{a5}) is integrable
if and only if its principle part is integrable.

\noindent
The principle part of (\ref{a5}) is obtained by ignoring the last
term (the term with $\chi^{i}_{j}$). Hence the proof of this corollary follows
directly by observing that the constraints on the constants
$\chi^{i}_{j}$ and $w^{i}_{j}$ listed in (\ref{c1}-\ref{c5}) are
independent of the constraints on the constants of
the principle part listed in (\ref{b4}-\ref{b11}). Before the application
of this corollary let us go back to the proposition 4 and ask the question
whether the KdV system (\ref{a5}) admits a recursion operator with
$w^{i}_{j}=0$.

\noindent
{\bf Corollary 2:} The KdV system (\ref{a5})  admits a recursion
operator of the principle part then the last term $\chi^{i}_{j}\,q^{j}_{,x}$
is a symmetry of the principle part.

\noindent
If $w^{i}_{j}=0$ the above equations (\ref{c1}-\ref{c5}) reduce to

\begin{eqnarray}
\chi^{k}_{l}\,a^{i}_{jk}-\chi^{i}_{k}\,a^{k}_{jl}-\chi^{i}_{k}\,c^{k}_{jl}
+\chi^{k}_{j}\,c^{i}_{kl}=0  ,\label{d1}\\
\chi^{k}_{l}\,s^{i}_{jk}-\chi^{i}_{k}\,s^{k}_{jl}=0  ,\label{d2} \\
\chi^{k}_{j}\,b^{i}_{k}-\chi^{i}_{k}\,b^{k}_{j}=0  ,\label{d3} \\
\chi^{k}_{l}\,c^{i}_{jk}-\chi^{i}_{k}\,c^{k}_{jl}=0  \label{d4}
\end{eqnarray}

\noindent
In order that the term $\chi^{i}_{j}\,q^{j}_{,x}$ to be a symmetry of
the principle part the constants $\chi^{i}_{j}$ are subject to satisfy
the following equations

\begin{eqnarray}
\chi^{k}_{l}\,s^{i}_{jk}-\chi^{i}_{k}\,s^{k}_{jl}=0  ,\label{e1} \\
\chi^{k}_{j}\,b^{i}_{k}-\chi^{i}_{k}\,b^{k}_{j}=0  ,\label{e2} \\
\chi^{k}_{l}\,K^{i}_{jk}+\chi^{k}_{j}\,K^{i}_{lk}=0  ,\label{e3}
\end{eqnarray}

\noindent
These equations simply follow from the set of equations (\ref{d1}-\ref{d4})
and hence the quantities
$\sigma^{i}=\chi^{i}_{j}\,q^{j}_{,x}$ are symmetries of the principle part.

By the application of corollary 1,
the full classification of the system (\ref{a5}) with the recursion
operator (\ref{a6}) such that $q^{i}_{,t}$ (i.e., the system of equations
themselves) belong to the symmetries of this system is possible. To each
subclass given in the previous sections there exists a Fokas-Liu extension
such that $w^{i}_{j}=\chi^{i}_{j}$ with the following constraints

\begin{eqnarray}
\chi^{i}_{k}\,c^{k}_{jl}-\chi^{k}_{l}\,c^{i}_{jk}=0 , \label{x1} \\
\chi^{k}_{l}\, a^{i}_{jk}-\chi^{k}_{j}\,a^{i}_{lk}-\chi^{i}_{k}\,(
a^{k}_{jl}-a^{k}_{lj})=0. \label{x2}
\end{eqnarray}

The above constraints are identically satisfied for the class
($det(b^{i}_{j}) \not =0$ , symmetrical case) when
$\chi^{i}_{j}= \alpha\, \delta^{i}_{j}+ \beta\, b^{i}_{j}$.
Hence the Fokas-Liu extension
of non-degenerate-symmetrical case is straightforward with this choice of
$\chi^{i}_{j}$. Here $\alpha$ and $\beta$ are arbitrary constants.

For the degenerate case the set of equations (\ref{x1} - \ref{x2}) must
solved for a given principle part , $b^{i}_{j}$ and $a^{i}_{jk}$.
Recently  a system of integrable KdV system  with $N=2$
has been introduced by Fokas and Liu \cite{fok}. This system is a
nice example for the application of corollary 1. We shall
give this system in its original form first and then simplify

\begin{eqnarray}
u_{,t}+v_{,x}+(3\, \beta_{1}+2\, \beta_{4})\, \beta_{3}\,u\, u_{,x}+
(2+\beta_{1}\, \beta_{4})\, \beta_{3}\, (uv)_{,x}+ \nonumber \\
\beta_{1}\, \beta_{3}\,v\, v_{,x}
+(\beta_{1}+ \beta_{4})\, \beta_{2}\,u_{,xxx}
+(1+\beta_{1}\, \beta_{4})\, \beta_{2}\, v_{,xxx}=0 ,\\
v_{,t}+u_{,x}+(2+3\, \beta_{1}\, \beta_{4})\, \beta_{3}\,v\,v_{,x}+
(\beta_{1}+2\, \beta_{4})\, \beta_{3}\, (uv)_{,x}+ \nonumber \\
\beta_{1}\, \beta_{3}\,\beta_{4}\,u\, u_{,x}+
(\beta_{1}+\beta_{4})\, \beta_{2}\,\beta_{4}\,u_{,xxx}
+(1+\beta_{1}\, \beta_{4})\, \beta_{2}\,\beta_{4} v_{,xxx}=0 .
\end{eqnarray}

\noindent
where $\beta_{1}$, $\beta_{2}$ , $\beta_{3}$ , and $\beta_{4}$ are
arbitrary constants. The recursion operator of this system is given
in \cite{fok}. Consider now a linear transformation

\begin{equation}
u=m_{1}\, r+n_{1}\,s~~,~~v=m_{2}\,r+n_{2}\,s
\end{equation}

\noindent
where $m_{1}$ , $m_{2}$ , $n_{1}$ , $n_{2}$ are constants and $s$ , $r$
are new dynamical variables, $q^{i}=(s,r)$.
Choosing these constants properly, the Fokas-Liu system reduces to a more
simpler form

\begin{eqnarray}
r_{,t}=(rs)_{,x}+\alpha_{1}\,r_{,x}+\alpha_{2}\,s_{x} , \nonumber \\
s_{,t}=\gamma_{1}\,s_{,xxx}+\gamma_{2}\,r\,r_{,x}+3\,s\,s_{,x}
+\alpha_{3}\,r_{,x}+\alpha_{4}\,s_{,x}, \label{f1}
\end{eqnarray}

\noindent
where we are not giving the coefficients $\alpha_{1},\alpha_{2}
,\alpha_{3}$ and $\alpha_{4}$ in terms of the parameters of the
original equation given above, because these expressions
are quite lengthy. The only condition on the parameters
$\alpha_{i}$  is given by $\alpha_{3}=\gamma_{2}\, \alpha_{2}$.
this guarantees the integrability of the above system
(\ref{f1}). On the other hand the transformation parameters
are given by

\begin{eqnarray}
m_{2}=-{\beta_{1}+ \beta_{4} \over 1+ \beta_{1}\, \beta_{4}}\,m_{1}~~,~~
n_{2}=\beta_{4}\,n_{1},\\
n_{1}=-{1 \over \delta\, \beta_{3}}~~,~~\delta=\beta_{1}\,(1+\beta_{4}^2)+
2 \beta_{4}.
\end{eqnarray}

\noindent
The principle part of the Fokas-Liu system (\ref{f1}) is exactly
the Ito system given in (\ref{ito}) and hence the recursion operator is
the sum of the one given in (\ref{rec}) and $\chi^{i}_{j}$ which are
given by $\chi^{1}_{1}=\alpha_{4}$\, , \, $\chi^{1}_{2}=\alpha_{3}$ \, ,
\, $\chi^{2}_{1}=\alpha_{2}$ \, , \, $\chi^{2}_{2}=\alpha_{1}$. That is

\begin{equation}
R=
\: \left(
\begin{array}{cc}
\gamma_{1}\,D^{2}+2\,s +s_{x}D^{-1}+\alpha_{4}
 & \gamma_{2}\, r+\alpha_{3} \\
r+r_{x}D^{-1}+\alpha_{2} &
\alpha_{1}
\end{array} \; \; \right)\;.
\label{rec1}
\end{equation}

\noindent
Another example is given very recently \cite{BON} in very different context
for $N=2$.

\begin{eqnarray}
&&u_{,t}={1 \over 2}\,v_{,xxx}+2\,u\,v_{,x}+u_{,x}\,v ,\nonumber \\
&&v_{,t}=3\,v\,v_{,x}+2\,a\,u_{,x}. \label{e20}
\end{eqnarray}

\noindent
The principle part of this equations is transformable to the
Fuchssteiner system given in (\ref{b19}). Taking $\alpha_{1}=2,
\alpha_{3}=4$ and scaling $x$ and $t$ properly we get (without
loosing any generality)

\begin{eqnarray}
 &&r_{t}=2\,r_{xxx}+6 r r_{x} ,\nonumber \\
 &&s_{t}=2 s_{x} r+4 s r_{x}.\label{e21}
\end{eqnarray}

\noindent
The transformation between the principle part ($a=0$)
of (\ref{e20}) and (\ref{e21}) is simply given
by $u=m_{1}\,s+{1 \over 2}\,r$ , $v=2\,r$. Then the recursion
operator of the system (\ref{e20}) is given by

\begin{equation}
R=
\: \left(
\begin{array}{cc}
0
 & {1 \over 2}D^{2}+2 u+u_{x}D^{-1} \\
2 a &
2 v+v_{x}D^{-1}
\end{array} \; \; \right)\;.
\label{rec5}
\end{equation}

{\bf{V. Conclusion}} \\
We have given a classification of a system of KdV equations with respect to
the existence of a recursion operator. This is indeed a partial
classification. Although we have found all conditions for each subclassess
we have not presented them explicitly. We obtained
three distinct subclasses for all values of $N$ and gave the corresponding
recursion operators. We also gave an extension of such systems by adding
a linear term containing the first derivative of dynamical variables.
We called such systems as the the Fokas-Liu extensions. We proved that
these extended systems of KdV equations are also integrable if and only if
their principle parts are integrable. For $N=2$ , we have given all
subclassess explicitly. Among these the recursion operator of the KdV
coupled to the time evolution part of its Lax-pair seems to be new. Here we
would like to add that when $N=2$ recursion operators , including the
Fokas-Liu extensions, are hereditary \cite{fk1}.

Our classification crucially depends on the form of the recursion
operator. The recursion operators used in this work were assumed to
have degree two (highest degree of the operator $D$ in $R$) and non-locality
order one (highest degree of the operator $D^{-1}$ in $R$). The next
works in this program should be the study on the classification problems
with respect to the recursion operators with higher
degree and higher non-localities. For instance when $N=2$ , Hirota-Satsuma,
Boussinesq and Bogoyavlenskii coupled KdV equations admit recursion
operators with $m=4$ and $n=1$ \cite{kar}. Hence these equations
do not belong to our classification given in this work.

This work is partially supported by the Scientific and Technical
Research Council of Turkey (TUBITAK). M.G is a member Turkish
Academy of Sciences (TUBA).

\end{document}